\documentclass[11pt]{article} 

\usepackage{graphicx} 
\graphicspath{{H:/Pictures/}}

\title{\vspace{-1.5cm} Two-part models with stochastic processes for modelling longitudinal semicontinuous data: computationally efficient inference and modelling the overall marginal mean}
\date{}
\author{Sean Yiu* and Brian D. M. Tom\\
*Email-address: Sean.yiu@mrc-bsu.cam.ac.uk\\
MRC Biostatistics Unit, School of Clinical Medicine,\\
 University of Cambridge, Cambridge, CB2 0SR, UK}
\usepackage[title,titletoc,toc]{appendix}
\usepackage{amsmath}
\usepackage{amssymb}
\usepackage{fullpage}
\usepackage{xypic}
\usepackage{rotating}
\usepackage{afterpage}
\usepackage{hhline}
\usepackage{booktabs}
\usepackage{bm}
\usepackage{rotfloat}

\usepackage{tikz}
\usetikzlibrary{fit,positioning}

\begin{document}
\maketitle
\noindent \textbf{Abstract}\\
Several researchers have described two-part models with patient-specific stochastic processes for analysing longitudinal semicontinuous data. In theory, such models can offer greater flexibility than the standard two-part model with patient-specific random effects. However, in practice the high dimensional integrations involved in the marginal likelihood (i.e. integrated over the stochastic processes) significantly complicates model fitting. Thus non-standard computationally intensive procedures based on simulating the marginal likelihood have so far only been proposed. In this paper, we describe an efficient method of implementation by demonstrating how the high dimensional integrations involved in the marginal likelihood can be computed efficiently. Specifically, by using a property of the multivariate normal distribution and the standard marginal cumulative distribution function identity, we transform the marginal likelihood so that the high dimensional integrations are contained in the cumulative distribution function of a multivariate normal distribution, which can then be efficiently evaluated. Hence maximum likelihood estimation can be used to obtain parameter estimates and asymptotic standard errors (from the observed information matrix) of model parameters. We describe our proposed efficient implementation procedure for the standard two-part model parameterisation and when it is of interest to directly model the overall marginal mean. The methodology is applied on a psoriatic arthritis data set concerning functional disability.
\section{Introduction}
Semicontinuous data arise when the outcome is a mixture of true zeros and continuously distributed positive values.$^1$ Some examples in the literature have included average daily alcohol consumption,$^1$ hospital lengths of stay$^2$ and medical expenditures.$^{3,4}$ In these situations, and more generally, it is natural to view the outcome as a result of two processes, the first determines if the outcome is zero, and if not the second determines the positive value. Two-part models are therefore convenient for the analysis of semicontinuous data and have been used extensively. Recently, Smith {\it et al.}$^{3,4}$ considered the interesting notion of reparameterising the mean of the positive values in terms of the overall mean, which is arguably a more justified target of inference (see Tom {\it et al.}$^{5}$ and the references therein). We also consider this notion with respect to the overall marginal mean in our framework.\\
\indent Two-part marginal models and two-part mixed models have both been proposed for the analysis of longitudinal semicontinuous data. The first is motivated by obtaining population-based inference and have been constructed using generalized estimating equations.$^6$ The second is more convenient when patient-specific inference is of interest, and are constructed by incorporating correlated patient-specific random effects in both parts of the model.$^7$ This paper focuses on the two-part mixed modelling approach, although considerations are provided on how population-based inference can be obtained. \\
\indent In some situations, correlated patient-specific random effects models will not provide an adequate fit to the data. This may especially be the case when the lengths of follow-up are relatively long. Here it may be less plausible to assume that patients can only have consistently high or low outcomes throughout their entire follow-up. In terms of the correlation structure, it may not be reasonable to assume constant correlation between outcomes from the same patient regardless of their gap times (which is induced by patient-specific random effects). Flexible two-part models that allow for random changes in the trajectory through serially correlated stochastic processes may then be more plausible and these have been proposed in the literature. Albert and Shen$^8$ and Ghosh and Albert$^9$ proposed two-part mixed models that consisted of correlated Gaussian processes and random walks (in addition to correlated patient-specific random effects) respectively in both parts of the model. Albert and Shen$^8$ demonstrated, through their application and a simulation study, that overall conditional means may suffer from bias if serial correlation (which is not captured by patient-specific random effects) is present but ignored. It is also worth noting, both models incorporating stochastic processes provided considerable improvements of fit to their data.\\
\indent A main drawback of fitting models with stochastic processes is the computationally intensive nature of the model fitting procedure. The primary difficulty results from the following feature: if a patient has $m_i$ observations, then a model consisting of correlated stochastic processes in each part of the model will require $2m_i$ integrations to evaluate the marginal likelihood contribution from that patient (assuming, as is usual, the stochastic processes are realised at the observation times). For manageable values of $m_i$, Albert and Shen$^8$ and Ghosh and Albert$^9$ have developed methods based on a Monte Carlo Expectation Maximization algorithm and Markov chain Monte Carlo respectively to evaluate the marginal likelihood. Both of these procedures can be computationally intensive, with the former also requiring standard errors of parameter estimates to be computed by bootstrap. The primary aim of this paper is to demonstrate, using a property of the multivariate normal distribution and the standard marginal cumulative distribution function identity, how a marginal likelihood can be obtained in terms of the cumulative distribution function of a multivariate normal distribution. Implicitly, because it is possible to efficiently evaluate the cumulative distribution function of a multivariate normal distribution, maximum likelihood estimation can be used to obtain parameter estimates and (asymptotic) standard errors (from the observed information matrix) of model parameters.\\
\indent The rest of this paper is organised as follows. In Section 2, the motivating application concerning functional disability in psoriatic arthritis is introduced. Section 3 describes the flexible two-part modelling framework of Albert and Shen$^8$ and Ghosh and Albert$^9$ (including additional comments regarding implementation). Section 4 proposes an efficient maximum likelihood estimation procedure for the models in Section 3. Section 5 applies the methodology in Section 4 to the data described in Section 2. While retaining the flexibility of using stochastic processes models and the practicality of the proposed efficient implementation procedure, Section 6 extends the modelling framework of Section 3 to allow for the direct modelling of the overall marginal mean. Finally, concluding remarks are made in Section 7.
\section{Functional disability in psoriatic arthritis}
Psoriatic arthritis (PsA) is an inflammatory arthritis associated with the skin condition psoriasis. Because of both skin and joint involvement of the disease, PsA can result in patients having severe physical functional disability. The dominant measure of functional disability in PsA, as well as in many other disease areas,$^{10}$ is the self reported Health Assessment Questionnaire (HAQ). This produces an essentially continuous measure$^{11-14}$ between zero, representing no disability, and three, representing severe disability.\\
\indent The HAQ scores of 698 patients observed longitudinally at the University of Toronto PsA clinic was considered for this analysis. Figure 1 shows the frequencies of HAQ scores from these patients. From Figure 1, it is evident that a large proportion of zeros exist in this data set (1526/4811=0.32). The clumping at zero, together with the continuous distributed outcomes for the non-zero values, suggests that the HAQ score can be viewed as a semicontinuous outcome. Su {\it et al.}$^{12,13}$ considered two-part models with patient-specific random effects for analysing an earlier version of this PsA data set. In this paper, we relax the assumption of constant patient-specific random effects to patient-specific stochastic processes and consider the extent to which they improve understanding of the disability process. This includes making easy interpretable inference on the overall marginal mean HAQ scores, a concept that has not been considered before with stochastic processes models (see Section 6 for more details). On average, patients had 6.89 clinic visits (ranging from 2 to 20) with mean inter-visit and follow-up times of 1 year and 5 months (standard deviation (SD) of 1 year and 1 month) and 8 years and 3 months (SD of 5 years and 10 months) respectively.

\begin{figure}[h!]
\centering
\includegraphics[scale=0.5]{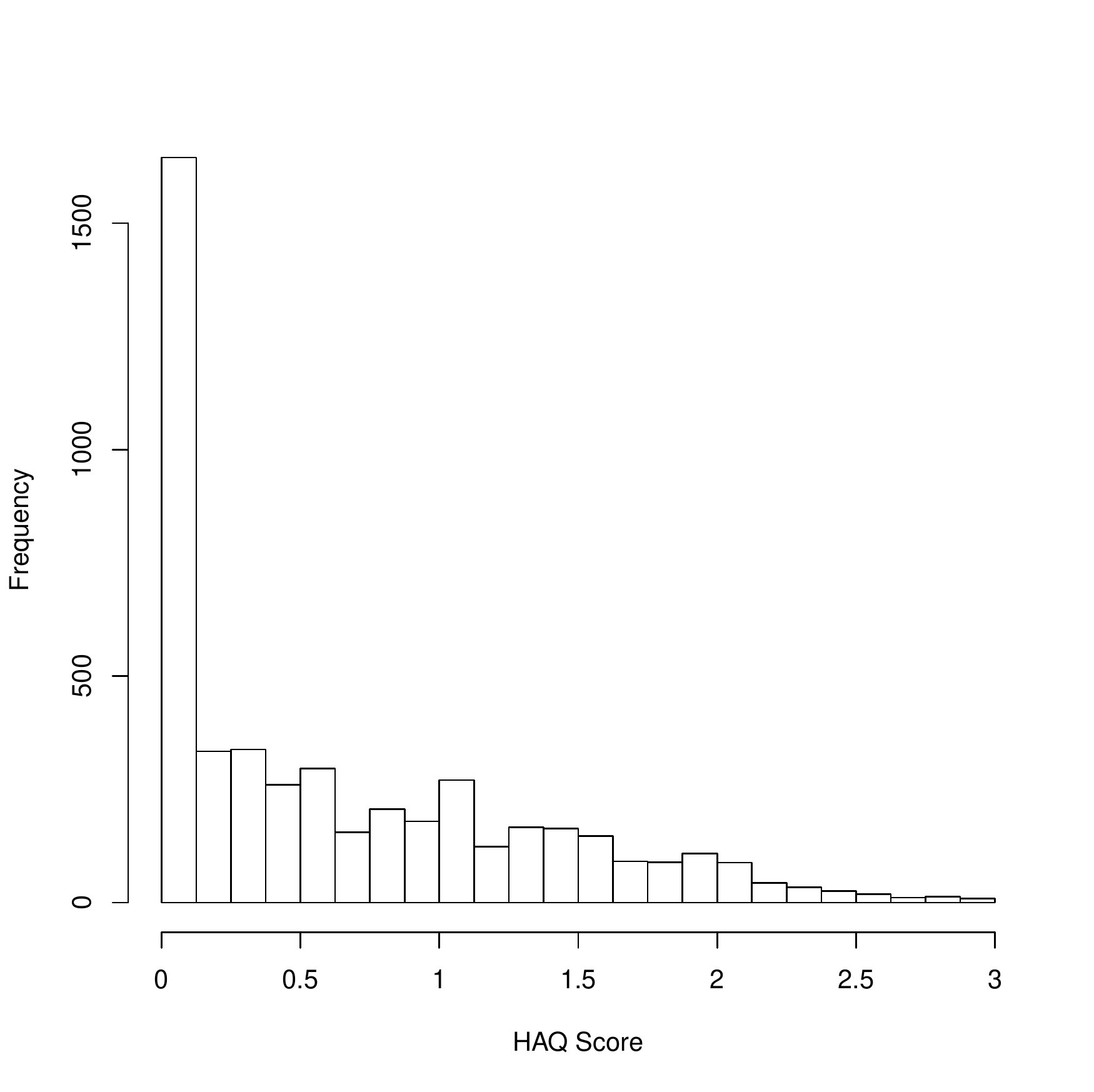}
\caption{Frequencies of HAQ scores in our data}
\end{figure}

%FIGURE 1 APPROXIMATELY HERE

%Figure 1: Frequencies of HAQ scores in our data.

\section{Model}
Let $Y_{ij}$ ($i=1,\ldots,N$) denote the semicontinuous response from patient $i$ at time $t_{ij}$ ($j=1,\ldots,m_i$), where $t_{ij}$ represents the time of the $j$th observation from patient $i$. Because of true zeros, it is natural to decompose the response into
\begin{displaymath}
U_{ij}=\left\{
     \begin{array}{lr}
     1& : Y_{ij}>0\\
     0& :Y_{ij}=0\\
     \end{array}
   \right.
\end{displaymath}
and $g(Y_{ij})|Y_{ij}>0$, where $g(\cdot)$ is a monotonic function such that $g(0)=0$ and $g(Y_{ij})|Y_{ij}>0$ is positive and approximately Gaussian with constant variance $\sigma^2$. For convenience, the model for $U_{ij}$ is referred to as the binary component, while the model for $g(Y_{ij})|Y_{ij}>0$ is referred to as the continuous component.\\
\indent We now describe the flexible modelling framework. Let $\bm{X}_{ij}$ and $\bm{Z}_{ij}$ be column vectors of covariates that influence the probability of $Y_{ij}>0$ and the mean of $g(Y_{ij})|Y_{ij}>0$ respectively. Then conditional on correlated patient-specific random effects $(B^r_i=b^r_i,C^r_i=c^r_i)$ and correlated stochastic processes $\{B^s_i(t_{ij})=b^s_{ij},C^s_i(t_{ij})=c^s_{ij}\}$, where the random effects are assumed independent of the stochastic processes, we model $U_{ij}$ as Bernoulli with response probability
\[\tag{1}\mathbb{P}(U_{ij}=1|b^s_{ij},b^r_i)=\Phi(\bm{X}^\top_{ij}\bm{\beta}+b^s_{ij}+b^r_i)\]
where $\Phi(\cdot)$ is the cumulative distribution function of a standard Gaussian distribution (i.e. probit model), and $[g(Y_{ij})|Y_{ij}>0;c^s_{ij},c^r_i]$ as Gaussian with mean $\bm{Z}^\top_{ij}\bm{\gamma}+c^s_{ij}+c^r_i$ and constant variance $\sigma^2$ (i.e. linear mixed effect model on  $g(Y_{ij})|Y_{ij}>0$). Here $\bm{\beta}$ and $\bm{\gamma}$ are column vectors of regression coefficients. The patient-specific random effects $(B^r_i,C^r_i)$ allow patients to have a consistently high or low probability of having disability and a consistently high or low mean for the non-zero HAQ scores across time. While the patient-specific stochastic processes $\{B^s_i(t_{i1})$,$\ldots$,$B^s_i(t_{im_i})$,$C^s_i(t_{i1}),\ldots$,$C^s_i(t_{im_i})\}$ can capture serial correlation and non-predictable changes in unobserved heterogeneity.$^9$\\
\indent We assume $\{B^r_i,C^r_i\}$ follows a bivariate normal distribution with mean vector zero and 
\[\tag{2}\text{Var}(B^r_i)=\sigma^2_b,~~~\text{Var}(C^r_i)=\sigma^2_c,~~~\text{Cov}(B^r_i,C^r_i)=\sigma_b\sigma_c\rho\]
where $\sigma^2_b$ and $\sigma^2_c$ are variance parameters and $\rho$ is the correlation between $B^r_i$ and $C^r_i$. Furthermore, we consider two classes of stochastic processes for $\{B^s_i(t),C^s_i(t)\}$ that are subsequently described. For convenience, define $B_i(t)=B^r_i+B^s_i(t)$ and $C_i(t)=C^r_i+C^s_i(t)$, i.e. the patient-specific random effects $B^r_i$ and $C^r_i$ are absorbed into the stochastic processes $B_i(t)$ and $C_i(t)$ respectively, and let the covariance matrix of $\{B_i(t_{i1})$,$\ldots$,$B_i(t_{im_i})$,$C_i(t_{i1}),\ldots$,$C_i(t_{im_i})\}$ be
\[\bm{\Sigma}_i=\bordermatrix{&\cr
                  &\bm{\Sigma}_{ib} &\bm{\Sigma}_{ibc}   \cr
           &\bm{\Sigma}_{ibc}&\bm{\Sigma}_{ic} \cr}.\]
\subsection{Correlated Gaussian processes}
The first and most general model that we consider is defined when $\{B^s_i(t),C^s_i(t)\}$ are correlated stationary Gaussian processes. That is the model proposed by Albert and Shen:$^8$
\begin{align*}\tag{3}
&\text{Cov}\{B^s_i(t_{ij}),B^s_i(t_{ik})\}=\sigma^2_{gb}\rho^{|t_{ij}-t_{ik}|}_{gb},~~~\text{Cov}\{C^s_i(t_{ij}),C^s_i(t_{ik})\}=\sigma^2_{gc}\rho^{|t_{ij}-t_{ik}|}_{gc},\\
&\text{Cov}\{B^s_i(t_{ij}),C^s_i(t_{ik})\}=\sigma_{gb}\sigma_{gc}\rho_g\rho^{|t_{ij}-t_{ik}|}_{gbc}
\end{align*}
where $\sigma^2_{gb}$ and $\sigma^2_{gc}$ are variance parameters, $\rho_{g}$ is the correlation between the Gaussian processes at the same time point, and $\rho_{gb}$, $\rho_{gc}$, $\rho_{gbc}$ are the degradation parameters governing the serial correlation within and between processes respectively. Following Albert and Shen,$^8$ the processes $B_i(t)$ and $C_i(t)$ are taken to be exchangeable Ornstein-Uhlenbeck (EOU) processes, and the model containing these processes is called the general model, i.e. (1-3). Some special cases of the general model are
\begin{itemize}
\item Shared EOU process model when $c_{ij}=\theta b_{ij}$,
\item Correlated OU processes model when $\sigma^2_{b}\equiv\sigma^2_{c}\equiv\rho\equiv 0$,
\item Shared OU process model when $c^s_{ij}=\theta b^s_{ij}$ and $\sigma^2_{b}\equiv\sigma^2_{c}\equiv\rho\equiv 0$,
\item Correlated random effects model when $\sigma^2_{gb}\equiv\sigma^2_{gc}\equiv\rho_g\equiv\rho_{gb}\equiv\rho_{gc}\equiv\rho_{gbc}\equiv 0$,
\item Shared random effect model when $c^r_{i}=\theta b^r_i$ and $\sigma^2_{gb}\equiv\sigma^2_{gc}\equiv\rho_g\equiv\rho_{gb}\equiv\rho_{gc}\equiv\rho_{gbc}\equiv 0$,
 \end{itemize}
where $\theta$ is a parameter to be estimated.
\subsubsection{Remarks on $\rho_{gb}$, $\rho_{gc}$ and $\rho_{gbc}$}
 Although the general model is very flexible, it will not always be mathematically valid. Let the covariance matrices $\bm{\Sigma}^s_{ib}$, $\bm{\Sigma}^s_{ic}$ and $\bm{\Sigma}^s_{ibc}$ have $(j,k)$th entry $\sigma^2_{gb}\rho^{|t_{ij}-t_{ik}|}_{gb}$, $\sigma^2_{gc}\rho^{|t_{ij}-t_{ik}|}_{gc}$ and  $\sigma_{gb}\sigma_{gc}\rho_g\rho^{|t_{ij}-t_{ik}|}_{gbc}$ respectively, i.e. described by (3). If $\rho_{gb}$, $\rho_{gc}$ and $\rho_{gbc}$ are unconstrained (as specified by Albert and Shen$^8$), the matrix $\bm{\Sigma}^s_i$ where
\[\bm{\Sigma}^s_i=\bordermatrix{&\cr
                  &\bm{\Sigma}^s_{ib} &\bm{\Sigma}^s_{ibc}   \cr
           &\bm{\Sigma}^s_{ibc}&\bm{\Sigma}^s_{ic} \cr}\]
will not in general be a valid covariance matrix since $\bm{\Sigma}^s_i$, although symmetric, is not constrained to be positive semi-definite and therefore $B^s_i(t)$ and $C^s_i(t)$ will not necessarily form a jointly Gaussian process. The primary difficulty results when $\rho_g$ (the correlation between $B^s_i(t)$ and $C^s_i(t)$ at each time $t$) is close to one because the processes $B^s_i(t)$ and $C^s_i(t)$ are similar and therefore it will not be plausible for them to degrade at vastly different rates (i.e. for $\rho_{gb}$, $\rho_{gc}$ and $\rho_{gbc}$ to be vastly different). A reasonable approximation in this situation would be to constrain the degradation and cross degradation parameters to be same, specifically $\rho_{gb}\equiv\rho_{gc}\equiv\rho_{gbc}\equiv\rho_{g1}$. This constraint would then enforce $\bm{\Sigma}^s_i$ to be a valid covariance matrix since the Schur component $\bm{\Sigma}^s_{ib}-\bm{\Sigma}^s_{ibc}(\bm{\Sigma}^s_{ic})^{-1}\bm{\Sigma}^s_{ibc}\equiv\sigma^2_{gb}(1-\rho^2_g)\bm{\Sigma}^s_{i}(\rho_{g1})$, where $\bm{\Sigma}^s_{i}(\rho_{g1})$ has $(j,k)$th entry $\rho^{|t_{ij}-t_{ik}|}_{g1}$, is constrained to be positive semi-definite. The resulting correlation structure would then be
\begin{align*}\tag{4}
&\text{Cor}\{B^s_i(t_{ij}),B^s_i(t_{ik})\}=\text{Cor}\{C^s_i(t_{ij}),C^s_i(t_{ik})\}=\rho^{|t_{ij}-t_{ik}|}_{g1},\\
&\text{Cor}\{B^s_i(t_{ij}),C^s_i(t_{ik})\}=\rho_g\rho^{|t_{ij}-t_{ik}|}_{g1}.
\end{align*}
In the motivating application $\rho_g$ was estimated close to one. Slight deviations from the correlation structure described by (4) (for example $\rho_{gb}\equiv\rho_{gc}\equiv\rho_{g1}$ and $\rho_{gbc}=\rho_{g1}\delta$ where $\delta\in (0,1)$) resulted in non-positive semi-definite matrices for various $\bm{\Sigma}_i$, and therefore the model fitting procedure was problematic. Note that a further simplification would be to constrain $\rho_g=1$ (in addition to $\rho_{g1}=\rho_{gb}$), this would result in the shared EOU process model. If however $\rho_g$ takes a smaller value, and therefore the two Gaussian processes are less correlated, it would then be more plausible for the Gaussian processes to degrade at different rates. Hence having unconstrained degradations parameters will likely be less problematic.\\
\indent For completeness, note that 
\begin{align*}\tag{5}
\{\bm{\Sigma}_{ib}\}_{jk}&=\sigma^2_b+\sigma^2_{gb}\rho^{|t_{ij}-t_{ik}|}_{gb},\\
\{\bm{\Sigma}_{ic}\}_{jk}&=\sigma^2_c+\sigma^2_{gc}\rho^{|t_{ij}-t_{ik}|}_{gc},\\
\{\bm{\Sigma}_{ibc}\}_{jk}&=\sigma_{b}\sigma_{c}\rho+\sigma_{gb}\sigma_{gc}\rho_g\rho^{|t_{ij}-t_{ik}|}_{gbc}.
\end{align*}

\subsection{Correlated random walks}
The second model structure that we consider is defined when $\{B^s_i(t),C^s_i(t)\}$ are correlated continuous-time random walks. That is the model proposed by Ghosh and Albert.$^9$ Specifically, define sequentially $\{B_i(t_{ij+1}),C_i(t_{ij+1})\}|\{B_i(t_{ij})=b_{ij},C_i(t_{ij})=c_{ij}\}$ to be bivariate normal with mean $(b_{ij},c_{ij})$ and covariance matrix 
\[\bordermatrix{&\cr
                  &\sigma^2_{wb}(t_{ij+1}-t_{ij}) &\sigma_{wb}\sigma_{wc}\rho_w(t_{ij+1}-t_{ij})   \cr
           &\sigma_{wb}\sigma_{wc}\rho_w(t_{ij+1}-t_{ij})&\sigma^2_{wc}(t_{ij+1}-t_{ij}) \cr}.\]
In addition $(b_{i1},c_{i1})=(b^r_{i},c^r_i)$ are initiated at realisations of the patient-specific random effects. Here $\sigma^2_{wb}$, $\sigma^2_{wc}$ and $\rho_{w}$ are variance and correlation parameters that quantify serial correlation (both within and across processes). This model will be denoted by a correlated random walks (CRW) model and it contains as special cases.
\begin{itemize}
\item Shared random walk model when $c_{ij}=\theta b_{ij}$,
\item Correlated random effects model when $\sigma^2_{wb}\equiv\sigma^2_{wc}\equiv\rho_w\equiv 0$,
\item Shared random effect model when $c^r_{i}=\theta b^r_i$ and $\sigma^2_{wb}\equiv\sigma^2_{wc}\equiv\rho_w\equiv 0$.
 \end{itemize}
Although the CRW model is less flexible than the general model, it has the advantage, from its sequential construction, of always being well defined even when the parameters are unconstrained (apart from the usual constraint that correlation parameters have modulus less than or equal to unity). Moreover  
\begin{align*}\tag{6}
\{\bm{\Sigma}_{ib}\}_{jk}&=\sigma^2_b+\sigma^2_{wb}\{\text{min}(t_{ij},t_{ik})-t_{i1}\},\\
\{\bm{\Sigma}_{ic}\}_{jk}&=\sigma^2_c+\sigma^2_{wc}\{\text{min}(t_{ij},t_{ik})-t_{i1}\},\\
\{\bm{\Sigma}_{ibc}\}_{jk}&=\sigma_{b}\sigma_{c}\rho+\sigma_{wb}\sigma_{wc}\rho_w\{\text{min}(t_{ij},t_{ik})-t_{i1}\}.
\end{align*}

\section{Efficient maximum likelihood estimation procedure for stochastic processes models}
This section describes our efficient maximum likelihood estimation procedure for the flexible models described in Section 3. Firstly, in Section 4.1 we describe a generic likelihood function for all of the described models. The multivariate normal identity that can be used to evaluate certain multi-dimensional integrals in terms of a multivariate normal cumulative distribution function is introduced in Section 4.2. Finally, in Section 4.3 we outline how to apply the multivariate normal identity in Section 4.2 to the generic likelihood function in Section 4.1, thus culminating in a computationally efficient likelihood. For completeness, we also provide computational simplifications for correlated stochastic processes models in the appendix.
\subsection{Likelihoods}
For ease of exposition, we describe the likelihood contribution from patient $i$. The likelihood can then be obtained by taking the product of all likelihood contributions from each patient. Firstly, we consider models that contain two (correlated) stochastic processes. For these models, the likelihood contribution from patient $i$ is
\begin{align*}\tag{7}
L_i(\bm{\Theta})=&\int_{\bm{b_i}}\int_{\bm{c_i}}\left[\prod_{j=1}^{m_i}\Phi(\bm{X}^\top_{ij}\bm{\beta}+b_{ij})^{u_{ij}}\{1-\Phi(\bm{X}^\top_{ij}\bm{\beta}+b_{ij})\}^{1-u_{ij}}\right]\\
\times&\left[\prod_{j=1}^{m_i}\left[\frac{1}{\sqrt{2\pi\sigma^2}}\exp\left\{-\frac{(g(y_{ij})-\bm{Z}^\top_{ij}\bm{\gamma}-c_{ij})^2}{2\sigma^2}\right\}\right]^{u_{ij}}\right]\phi^{(2m_i)}(\bm{b}_i,\bm{c}_i;\bm{0},\bm{\Sigma}_i)\,d\bm{b}_i\,d\bm{c}_i
\end{align*}
where $\bm{\Theta}$ is a vector representing all of the unknown parameters, $\bm{b}_i=(b_{i1},\ldots,b_{im_i})^\top$ and $\bm{c}_i=(c_{i1},\ldots,c_{im_i})^\top$, $\phi^{(m)}(.;\bm{\mu},\bm{\Sigma})$ is an $m$ dimensional multivariate normal density with mean vector $\bm{\mu}$ and covariance matrix $\bm{\Sigma}$, and $\bm{\Sigma}_i$ is defined by either (5) or (6). Similarly, for models containing a single stochastic process (i.e. shared process models), the likelihood contribution from patient $i$ is
\begin{align*}\tag{8}
L_i(\bm{\Theta})=&\int_{\bm{b_i}}\left[\prod_{j=1}^{m_i}\Phi(\bm{X}^\top_{ij}\bm{\beta}+b_{ij})^{u_{ij}}\{1-\Phi(\bm{X}^\top_{ij}\bm{\beta}+b_{ij})\}^{1-u_{ij}}\right]\\
\times&\left[\prod_{j=1}^{m_i}\left[\frac{1}{\sqrt{2\pi\sigma^2}}\exp\left\{-\frac{(g(y_{ij})-\bm{Z}^\top_{ij}\bm{\gamma}-\theta b_{ij})^2}{2\sigma^2}\right\}\right]^{u_{ij}}\right]\phi^{(m_i)}(\bm{b}_i;\bm{0},\bm{\Sigma}_{ib})\,d\bm{b}_i
\end{align*}
where $\bm{\Sigma}_{ib}$ can again be obtained from (5) or (6). We now define our generic likelihood contribution from patient $i$ which encompasses all of the described  models.  Throughout we apply the following notation: $\bm{0}$ and $\bm{1}$ are $m_i\times 1$ vectors with all entries being zero and one respectively, $\text{diag}(\bm{v})$ is a matrix with diagonal elements $\bm{v}$ and zero otherwise, and $\bm{I}_{d}$ is a $d\times d$ identity matrix. We also follow the convention that binary operations with a scalar and vector or matrix argument and unary operations with a vector argument are performed element-wise. In matrix form, we have 
\begin{align*}\tag{9}
L_i(\bm{\Theta})=&\int_{\bm{l}_i}\Phi^{(m_i)}\left(\bm{A}_{i1}\bm{\mu}_{ib}+\bm{A}_{i2}\bm{l}_i;\bm{0},\bm{I}_{m_i}\right)\left(\frac{1}{\sqrt{2\pi\sigma^2}}\right)^{\sum_{j=1}^{m_i}u_{ij}}\\
\times&\exp\left\{-\frac{(g(\bm{y}_i)-\bm{A}_{i3}\bm{\mu}_{ic}-\bm{A}_{i4}\bm{l}_i)^\top(g(\bm{y}_i)-\bm{A}_{i3}\bm{\mu}_{ic}-\bm{A}_{i4}\bm{l}_i)}{2\sigma^2}\right\}\phi^{\{\text{dim}(\bm{l}_i)\}}(\bm{l}_i;\bm{0},\bm{\Sigma}_{il})\,d\bm{l}_i
\end{align*}
where $\bm{y}_i=(y_{i1},\ldots,y_{im_i})^\top$, $\bm{\mu}_{ib}=\bm{X}_i\bm{\beta}$, $\bm{\mu}_{ic}=\bm{Z}_i\bm{\gamma}$, $\bm{X}_i=(\bm{X}_{i1},\ldots,\bm{X}_{im_i})^\top$, $\bm{Z}_i=(\bm{Z}_{i1},\ldots,\bm{Z}_{im_i})^\top$, and $\bm{A}_{i1}=\text{diag}(2\bm{u}_i-{\bm{1}})$, $\bm{A}_{i3}=\text{diag}(\bm{u}_i)$ are $m_i\times m_i$ matrices with $\bm{u}_i=(u_{i1},\ldots,u_{im_i})^\top$. Here $\Phi^{(d)}(.;0,\bm{\Sigma})$ represents the distribution function of $\phi^{(d)}(.;\bm{0},\bm{\Sigma})$ and $\bm{l}_i$ is a (to be specified) column vector of random effects. Note that (9) has resulted from repeated application of the identity $1-\Phi(x)=\Phi(-x)$.\\
\indent The likelihood contribution from patient $i$, $L_i(\bm{\Theta})$, is then obtained by specifying the vector of random effects $\bm{l}_i$ and its covariance matrix $\bm{\Sigma}_{il}$ together with the $m_i \times \text{dim}(\bm{l}_i)$ matrices $\bm{A}_{i2}$ and $\bm{A}_{i4}$ which describe how the random effects act on the binary and continuous components of the model. For (7), $\bm{l}_i=(\bm{b}_i,\bm{c}_i)$, $\bm{A}_{i2}=(\text{diag}(2\bm{u}_i-{\bm{1}}),\text{diag}({\bm{0}}))$ and $\bm{A}_{i4}=(\text{diag}({\bm{0}}),\text{diag}(\bm{u}_i))$. While for (8), $\bm{l}_i=\bm{b}_i$, $\bm{A}_{i2}=\text{diag}(2\bm{u}_i-{\bf{1}})$ and $\bm{A}_{i4}=\text{diag}(\theta \bm{u}_i)$. Similarly, for the correlated random effects model, $\bm{l}_i=(b_i,c_i)$, $\bm{A}_{i2}=(2\bm{u_i}-{\bf{1}},\bm{0})$ and $\bm{A}_{i4}=(\bm{0},\bm{u}_i)$, and for the shared random effect model, $l_i=b_i$, $\bm{A}_{i2}=2\bm{u}_i-{\bf{1}}$ and $\bm{A}_{i4}=\theta \bm{u}_i$.

\subsection{Multivariate normal identity}
In order to evaluate the likelihood described by (9), we derive a multivariate normal identity that makes use of a property of the multivariate normal distribution and the standard marginal cumulative distribution function identity. Firstly, suppose that $\bm{\omega}=(\bm{\omega}_1,\bm{\omega}_2)^\top$ follows a multivariate normal distribution with mean vector $(\bm{0},\bm{\eta})^\top$ where $\bm{\omega}_1$ and $\bm{0}$ are $k_1 \times 1$ vectors and $\bm{\omega}_2$ and $\bm{\eta}$ are $k_2 \times 1$ vectors respectively. Furthermore, suppose that the covariance matrix of $\bm{\omega}$ is the $(k_1+k_2) \times (k_1+k_2)$ matrix $\bm{\Sigma}$ where the first $k_1$ rows of $\bm{\Sigma}$ is the $k_1 \times (k_1+k_2)$ matrix $(\bm{\Sigma}_{22},\bm{\Sigma}^\top_{12})$ and the remaining $k_2$ rows of $\bm{\Sigma}$ is the $k_2 \times (k_1+k_2)$ matrix $(\bm{\Sigma}_{12},\bm{\Sigma}_{11})$ respectively. It is a well known result that $\phi^{(k_1+k_2)}(\bm{\omega};(\bm{0},\bm{\eta})^\top,\bm{\Sigma})=\phi^{(k_1)}(\bm{\omega}_1;\bm{\Sigma}^\top_{12}\bm{\Sigma}^{-1}_{11}(\bm{\omega}_2-\bm{\eta}),\bm{\Sigma}_{22}-\bm{\Sigma}^\top_{12}\bm{\Sigma}^{-1}_{11}\bm{\Sigma}_{12})$$\phi^{(k_2)}(\bm{\omega}_2;\bm{\eta},\bm{\Sigma}_{11})$ where the right-hand side is the product of the conditional density of $\bm{\omega_1}|\bm{\omega}_2$ and the marginal density of $\bm{\omega_2}$. By applying the standard marginal cumulative distribution function identity $F_{\bm{\omega}_1}(\bm{\omega}_1)=\int_{\bm{\omega}_2}F_{\bm{\omega}_1|\bm{\omega}_2}(\bm{\omega}_1|\bm{\omega}_2)f_{\bm{\omega_2}}(\bm{\omega}_2)\,d\bm{\omega}_2$ where the integrand is based on the right-hand side of the above result, we obtain the multivariate normal identity:  
\[\tag{10}\Phi^{(k_1)}(\bm{\omega}_1;\bm{0},\bm{\Sigma}_{22})=\int_{\bm{\omega}_2}\Phi^{(k_1)}\left(\bm{\omega}_1-\bm{\Sigma}^\top_{12}\bm{\Sigma}^{-1}_{11}(\bm{\omega}_2-\bm{\eta});\bm{0},\bm{\Sigma}_{22}-\bm{\Sigma}^\top_{12}\bm{\Sigma}^{-1}_{11}\bm{\Sigma}_{12}\right)\phi^{(k_2)}\left(\bm{\omega}_2;\bm{\eta},\bm{\Sigma}_{11}\right)\,d\bm{\omega}_2\]
by noting that the marginal distribution of $\bm{\omega}_1$ is multivariate normal with mean vector $\bm{0}$ and covariance matrix $\Sigma_{22}$.\\
\indent Returning to the application, the general idea is to rearrange (9) to take the form of the right-hand side of (10), and then to use (10) to compute the integrations over the random effects in terms of an $m_i$ dimensional normal cumulative distribution function. Because there exists efficient implementations of the multivariate normal cumulative distribution function, this approach will allow for the efficient computation of the generic likelihood. We note that Barrett {\it et al.}$^{15}$ used (10) to obtain computationally efficient likelihoods of flexible models that jointly consider longitudinal and time to event outcomes. Equation (10) also arises frequently in results concerning the multivariate skew normal distribution$^{16-19}$.
\subsection{Re-expressing the likelihoods}
This section demonstrates how (9) (the likelihood contribution from patient $i$) can be re-expressed. We firstly consider the integrand terms resulting from the continuous component and random effects. That is,
\[\tag{11}\left(\frac{1}{\sqrt{2\pi\sigma^2}}\right)^{\sum_{j=1}^{m_i}u_{ij}}\exp\left\{-\frac{(g(\bm{y}_i)-\bm{A}_{i3}\bm{\mu}_{ic}-\bm{A}_{i4}\bm{l}_i)^\top(g(\bm{y}_i)-A_{i3}\bm{\mu}_{ic}-\bm{A}_{i4}\bm{l}_i)}{2\sigma^2}\right\}\phi^{\{\text{dim}(\bm{l}_i)\}}(\bm{l}_i;\bm{0},\bm{\Sigma}_{il}).\]
By completing the square in $\bm{l}_i$ (see the appendix for more details), (11) can be rearranged as
\[\tag{12}L_{i1}\phi^{\{\text{dim}(\bm{l}_i)\}}(\bm{l}_i;\bm{h}_i,\bm{H}^{-1}_i)\]
where 
\begin{align*}\tag{13}
\bm{H}_i=&\bm{A}^\top_{i4}\bm{A}_{i4}/\sigma^2+(\bm{\Sigma}_{il})^{-1},\\
\bm{h}_i=&\bm{H}^{-1}_i\bm{A}^\top_{i4}(g(\bm{y}_i)-\bm{A}_{i3}\bm{\mu}_{ic})/\sigma^2
\end{align*}
and
\[\tag{14}
L_{i1}=\left(\frac{1}{\sqrt{2\pi\sigma^2}}\right)^{\sum_{j=1}^{m_i}u_{ij}}\frac{1}{|\bm{\Sigma}_{il} \bm{H}_i|^{1/2}}\exp\left\{-\frac{(g(\bm{y}_i)-\bm{A}_{i3}\bm{\mu}_{ic})^\top(g(\bm{y}_i)-\bm{A}_{i3}\bm{\mu}_{ic})}{2\sigma^2}+\frac{\bm{h}^\top_i\bm{H}_i\bm{h}_i}{2}\right\}
\]
is independent of $\bm{l}_i$. Substituting (12) into (9), we consider the integral (ignoring $L_{i1}$):
\[\tag{15}\int_{\bm{l}_i}\Phi^{(m_i)}\left(\bm{A}_{i1}\bm{\mu}_{ib}+\bm{A}_{i2}\bm{l}_i;\bm{0},\bm{I}_{m_i}\right)\phi^{\{\text{dim}(\bm{l}_i)\}}(\bm{l}_i;\bm{h}_i,\bm{H}^{-1}_i)\,d\bm{l}_i.\]
We can re-express the argument and covariance matrix of the multivariate normal distribution function in (15) as
\begin{align*}\tag{16}
\bm{A}_{i1}\bm{\mu}_{ib}+\bm{A}_{i2}\bm{l}_i&=\bm{A}_{i1}\bm{\mu}_{ib}+\bm{A}_{i2}\bm{h}_i-(-\bm{H}^{-1}_i\bm{A}^\top_{i2})^\top\bm{H}_i(\bm{l}_i-\bm{h}_i)\\
\bm{I}_{m_i}&=\bm{I}_{m_i}+\bm{A}_{i2}\bm{H}^{-1}_i\bm{A}^\top_{i2}-(-\bm{H}^{-1}_i\bm{A}^\top_{i2})^\top\bm{H}_i(-\bm{H}^{-1}_i\bm{A}^\top_{i2}).
\end{align*}
Therefore (15), after applying the multivariate normal identity (described by (10)), is equivalent to
\[\tag{17}\Phi^{(m_i)}\left(\bm{A}_{i1}\bm{\mu}_{ib}+\bm{A}_{i2}\bm{h}_i;\bm{0},\bm{I}_{m_i}+\bm{A}_{i2}\bm{H}^{-1}_i\bm{A}^\top_{i2}\right).\]
\indent Based on the above expressions, the likelihood contribution from patient $i$ can now be re-expressed as
\begin{align*}\tag{18}
L_i(\bm{\Theta})=&\Phi^{(m_i)}\left(\bm{A}_{i1}\bm{\mu}_{ib}+\bm{A}_{i2}\bm{h}_i;\bm{0},\bm{I}_{m_i}+\bm{A}_{i2}\bm{H}^{-1}_i\bm{A}^\top_{i2}\right)\left(\frac{1}{\sqrt{2\pi\sigma^2}}\right)^{\sum_{j=1}^{m_i}u_{ij}}\\
\times&\frac{1}{|\bm{\Sigma}_{il}\bm{H}_i|^{1/2}}\exp\left\{-\frac{(g(\bm{y}_i)-\bm{A}_{i3}\bm{\mu}_{ic})^\top(g(\bm{y}_i)-\bm{A}_{i3}\bm{\mu}_{ic})}{2\sigma^2}+\frac{\bm{h}^\top_i\bm{H}_i\bm{h}_i}{2}\right\}
\end{align*}
where 
\begin{align*}\tag{19}
\bm{A}_{i1}&=\text{diag}(2\bm{u}_i-{\bf{1}})\\
\bm{A}_{i3}&=\text{diag}(\bm{u}_i)\\
\bm{H}_i&=\bm{A}^\top_{i4}\bm{A}_{i4}/\sigma^2+(\bm{\Sigma}_{il})^{-1}\\
\bm{h}_i&=\bm{H}^{-1}_i\bm{A}^\top_{i4}(g(\bm{y}_i)-\bm{A}_{i3}\bm{\mu}_{ic})/\sigma^2
\end{align*}
and $\bm{\Sigma}_{il}=\bm{\Sigma}_i$ or $\bm{\Sigma}_{ib}$ with $\bm{A}_{i2}$ and $\bm{A}_{i4}$ defined by the specified model.\\ 

\indent From (18-19), it is now evident that evaluating the integrations involved in $L_i(\bm{\Theta})$ reduces to computing the cumulative distribution function of a multivariate normal distribution. This can be performed efficiently, for example by using the \verb R $^{20}$ package \verb mnormt $^{21}$. The model fitting procedure is then completed by maximizing the log-likelihood, for example by using the BFGS$^{22}$ optimization technique, to obtain parameter estimates and asymptotic standard errors (from the observed Fisher information matrix) of model parameters.
\section{Application: patient-specific inference}
Using the estimation procedure described in Section 4, we demonstrate how patient-specific inference on the probability of being disabled and the transformed mean HAQ score conditional on disability can be obtained. Specifically, how a unit change in covariate values impacts these quantities for any specific patient. We consider the covariate effects of the number of clinically damaged joints (time-dependent), the number of actively inflamed joints (time-dependent), sex (coded as 1 for males and 0 for females), arthritis duration in years (time-dependent), and age at onset of arthritis in years  (standardise). Following Su {\it et al.},$^{12,13}$ no transformation was applied to the non-zero HAQ scores, i.e. $g(y)=y$.\\
\indent Initially models with two stochastic processes were fitted to the HAQ data. This resulted in large estimated correlation parameters between the random effects (i.e. $\rho\approx 1$) and stochastic processes for both the correlated Gaussian processes and random walks cases (i.e. $\rho_{g}$ and $\rho_{w}\approx1$). These results therefore suggested a single stochastic process would be sufficient for describing the data. The shared EOU model was then fitted. However the analysis provided evidence for model over-parameterisation as $\hat{\sigma}^2$ appeared to converge at virtually zero and a positive-definite observed Fisher information matrix could not be attained (even when a considerably smaller tolerance level than the default was specified for the computation of multivariate normal probabilities). We therefore considered the shared random walk and OU process models, and for comparative purposes, the shared random effect model. The models containing stochastic processes were fitted using the likelihood described by (18-19), while the shared random effect model was fitted using numerical integration (since only a single integration per patient is required). The same parameter estimates for the shared random effect model were obtained when (18-19) were used in the model fitting procedure.\\
\indent Table 1 presents the results of the fitted models. Across the models, the covariate effects on the mean conditional on disability are seen to be relatively similar as the confidence intervals generally overlap. In addition, the models are in agreement with regard to the association of each covariate apart from arthritis duration. Arthritis duration is statistically significant in the shared random effect model but is not statistically significant in the models that incorporate stochastic processes. It is interesting to note that there are strong agreements regarding the covariate effect of the number of active joints (similar parameter estimates across models and relatively narrow confidence intervals). The models indicate an additional actively inflamed joint will increase the mean HAQ score conditional on disability by approximately 0.21 for any specific patient. For the binary component, the covariate effects are again seen to be relatively similar due to the overlapping confidence intervals. Their interpretation through the direction of association and statistical significance are also consistent across models. The covariate effects from the shared random effect model does however consistently demonstrate attenuation to the null when compared to the other models with stochastic processes.\\
\indent A generalized likelihood ratio test of $\sigma^2_{wb}=0$ and $\rho_{gb}=1$ produced $p$ values of $<0.001$ therefore suggesting preference towards the shared random walk and OU process models respectively when compared to the shared random effect model. Since the shared random walk and OU process models contain the same number of parameters, information criteria, such as AIC, would indicate (weakly) that the shared random walk model is preferable. It is also worth noting that the heterogeneity parameter in the binary component (i.e. $\sigma^2_b$ or $\sigma^2_{gb}$) is significantly lower in the shared random effect model. For this model, this parameter governs both the heterogeneity and correlation due to repeated measurements and therefore in light of greater unaccounted heterogeneity (compared to the models with stochastic processes), less correlation is expected.$^{23}$ In the continuous component, where $\sigma^2$ also accounts for heterogeneity, a smaller difference between the heterogeneity parameters (i.e. $\theta^2\sigma^2_b$ or $\theta^2\sigma^2_{gb}$) is seen; in the order of the models displayed in the table (from right to left), the heterogeneity parameters are 0.24, 0.36 and 0.25 respectively.   
\begin{sidewaystable}
\caption{Table displaying patient-specific effects and corresponding $95\%$ Wald intervals on the probability of being disabled and the mean HAQ score conditional on disability. *denotes the standardised version of the covariate.}
\begin{center}
\begin{tabular}{l*{6}{c}c}
\toprule
&Shared random walk&Shared OU process&Shared random effect\\
\midrule
Binary component &&\\
\midrule
Damaged joints &0.031 (0.013, 0.049) &0.04 (0.021, 0.059) &0.012 (0.002, 0.022)\\
Active joints&0.16 (0.13, 0.18) &0.17 (0.15, 0.2) &0.15 (0.13, 0.16) \\
Sex &-1.72 (-1.9, -1.54) &-2.17 (-2.92, -1.42) &-1.34 (-1.65, -1.02) \\
Arthritis duration &0.042 (0.025, 0.058) &0.044 (0.022, 0.065)&0.034 (0.023, 0.044) \\
Age at arthritis onset* &0.56 (0.45, 0.66)&0.68 (0.48, 0.88) &0.45 (0.3, 0.6)\\
Intercept &1.77 (1.62, 1.92)&1.97 (1.07, 2.87)&1.11 (0.81, 1.41) \\
\midrule
Continuous component&&\\
\midrule
Damaged joints &0.0065 (0.0032, 0.0097)&0.0078 (0.0046, 0.011)&0.0033 (0.00086, 0.0058)\\
Active joints  &0.021 (0.019, 0.023) &0.021 (0.019, 0.023)&0.02 (0.018, 0.022)\\
Sex &-0.29 (-0.34, -0.24)&-0.35 (-0.46, -0.23) &-0.29 (-0.37, -0.21)\\
Arthritis duration &0.0025 (-0.0008, 0.0058)&0.0035 (-0.001, 0.0081)&0.0067 (0.0041, 0.0093) \\
Age at arthritis onset* &0.076 (0.048, 0.1)&0.093 (0.046, 0.14) &0.086 (0.048, 0.12)\\
Intercept &0.62 (0.57, 0.66) &0.59 (0.44, 0.74) &0.63 (0.56, 0.7) \\
\midrule
$\theta$&0.2 (0.19, 0.22)&0.19 (0.17, 0.21) &0.27 (0.24, 0.3)\\
$\sigma^2$ &0.074 (0.069, 0.079) &0.066 (0.06, 0.072) &0.12 (0.11, 0.12)\\
$\sigma^2_b$ &6.3 (5.64, 7.04) & &3.29 (2.64, 4.1)\\
$\sigma^2_{gb}$  & &10.11 (8.03, 12.74) &\\
$\sigma^2_{wb}$ &0.58 (0.52, 0.65) & &\\
$\rho_{gb}$ & &0.95 (0.94, 0.96) &\\
Log-likelihood &-3279.08 &-3282.11 &-3500.48\\
\bottomrule
\end{tabular}
\end{center}
\end{sidewaystable}

\section{Modelling the overall marginal mean}
In many cases, it is of interest to obtain population-based inference in addition/as opposed to patient-specific inference. For example,  for strategic public health policy purposes it would be more clinically meaningful to obtain covariate effects on quantities of interest after averaging over all patients. Currently, the proposed models are parametrised to allow easily interpretable patient-specific covariate effects, those with $B_i(t_{ij})=b_{ij}$ and $C_i(t_{ij})=c_{ij}$, to act on the patient-specific mean of the transformed positive values (i.e. $\mathbb{E}[g(Y_{ij})|Y_{ij}>0,C_i(t_{ij})=c_{ij}]$) and the patient-specific probability of a having a positive value (i.e. $\mathbb{P}(U_{ij}=1|B_i(t_{ij})=b_{ij})$). However, under this parametrisation, it no longer becomes straightforward to obtain easily interpretable population-level covariate effects on the marginal mean of the transformed positive values (the mean of the transformed positive values after averaging over all $b_{ij}$ and $c_{ij}$, i.e. $\mathbb{E}[g(Y_{ij})|Y_{ij}>0]$) since it is a highly non-linear function of the linear predictors in the binary and continuous components.$^{5}$ Thus the effect of a single covariate is generally interpreted by fixing other covariates at certain values.$^8$ This problem remains even when population-level covariate effects on the overall marginal mean of the transformed values (i.e. $\mathbb{E}[g(Y_{ij})]$) are of primary interest, which has strongly been argued as an important target of inference;$^{24}$ it is estimated using data from the same patients over time (unlike $\mathbb{E}[g(Y_{ij})|Y_{ij}>0]$) and it is a measure of the undecomposed outcome. We reiterate that in considering the overall marginal mean of the transformed values as a target of inference, we assume that the monotonic transformation function is such that $g(0)=0$ and $g(Y_{ij})|Y_{ij}>0$ is positive and approximately Gaussian with constant variance $\sigma^2$.\\
\indent In order to obtain population-based inference on the overall marginal mean of the transformed values, Smith {\it et al.}$^4$ proposed the following model parameterisation
\begin{align*}\tag{20}
\mathbb{P}(U_{ij}=1|B^r_{i}=b^r_{i})&=g_1(\bm{X}^\top_{ij}\bm{\beta}+b^r_i)\\
\mathbb{E}[g(Y_{ij})|C^r_i=c^r_i]&=g_2(\bm{Z}^\top_{ij}\bm{\alpha}+c^r_i)
\end{align*}
where $g_1(\cdot)$ and $g_2(\cdot)$ are monotonic link functions and $B^r_i, C^r_i$ are, as before, zero mean bivariate normal patient-specific random effects. Recall that transformation and link functions differ in that transformation functions are applied prior to modelling. In their specific context, Smith {\it et al.}$^4$ considered the identity transformation for $g(\cdot)$ but allowed the positive values of $Y_{ij}$ to follow a log-skew-normal distribution. Under this parametrisation, for a suitably chosen link function such as $g^{-1}_2$ being the identity or log link, it is implicit that easily interpretable covariate effects on the overall marginal mean of $g(Y_{ij})$, $\bm{\alpha}$, can now be obtained. Smith {\it et al.}$^4$ implemented this model by using a Bayesian estimation approach with  
\[\mathbb{E}[g(Y_{ij})|Y_{ij}>0,B^r_{i}=b^r_i,C^r_{i}=c^r_{i}]=\frac{g_2(\bm{Z}^\top_{ij}\bm{\alpha}+c^r_i)}{g_1(\bm{X}^\top_{ij}\bm{\beta}+b^r_i)}\]
specified in the likelihoods defined by (7) or (8). Note that $\mathbb{E}[g(Y_{ij})|Y_{ij}>0,B^r_{i}=b^r_i,C^r_{i}=c^r_{i}]$ is no longer parametrised to be equivalent to a monotonic function of a linear predictor, as was specified before. While this approach for modelling the overall marginal mean is intuitive, it is clear that the multivariate normal identity in Section 4.2 can no longer be used to compute the integrations over the multi-dimensional random effects in the marginal likelihood. Thus, as mentioned in the introduction, implementation of such models can be computationally challenging, especially for our situation where it would be of interest to consider $b_{ij}$ and $c_{ij}$ (i.e. realisations of stochastic processes) instead of $b^r_i$ and $c^r_i$ (i.e. realisations of patient-specific random effects) in (20).\\ 
\indent We now propose another method which would allow easily interpretable covariate effects to act on the overall marginal mean of $g(Y_{ij})$. In contrast, this method facilitates the inclusion of stochastic processes because it retains the proposed efficient implementation procedure described in Section 4. To the best of our knowledge, there are no other methods in the literature that facilitates the practical implementation of stochastic processes models for directly modelling the overall marginal mean.\\
\indent We first begin by computing the overall marginal mean of $g(Y_{ij})$ when
\begin{align*}\tag{21}
\mathbb{P}(U_{ij}=1|B_i(t_{ij})=b_{ij})&=g_1(\bm{X}^\top_{ij}\bm{\beta}+b_{ij})\\
\mathbb{E}[g(Y_{ij})|Y_{ij}>0,C_i(t_{ij})=c_{ij}]&=\Delta_{ij}+c_{ij}\\
\end{align*}
where $\Delta_{ij}$ is a function of covariates (at the $j$th visit from patient $i$) and regression coefficients only. That is $c_{ij}$ is now assumed to act linearly on the mean of the transformed positive values and not on the overall mean of the transformed values as is the case in (20). For models with two processes, the overall marginal mean of $g(Y_{ij})$ is defined by
\[\mathbb{E}[g(Y_{ij})]=\int_b\int_c\Phi(\bm{X}^\top_{ij}\bm{\beta}+b_{ij})(\Delta_{ij}+c_{ij})\phi^{(2)}(b_{ij},c_{ij};\bm{0},\bm{\Sigma}_{ij})\,db_{ij}\,dc_{ij}\]
where 
\[\bm{\Sigma}_{ij}=\bordermatrix{&\cr
                  &\sigma^2_{bij} &   \sigma_{bij}\sigma_{cij}\rho_{bcij}  \cr
           & \sigma_{bij}\sigma_{cij}\rho_{bcij}&  \sigma^2_{cij}\cr}\]
and $\sigma^2_{bij}$, $\sigma^2_{cij}$ and $\rho_{bcij}$ are the variances and correlation of $B_i(t_{ij})$ and $C_i(t_{ij})$ respectively.
Similarly, the overall marginal mean for a shared process model is given by
\[\mathbb{E}[g(Y_{ij})]=\int_b\Phi(\bm{X}^\top_{ij}\bm{\beta}+b_{ij})(\Delta_{ij}+\theta b_{ij})\phi^{(1)}(b_{ij};0,\sigma^2_{bij})\,db_{ij}.\]
Conveniently, these integrals can be computed analytically and this results in 
\[\Phi\left(\frac{\bm{X}^\top_{ij}\bm{\beta}}{\sqrt{1+\sigma^2_{bij}}}\right)\Delta_{ij}+\frac{\sigma_{bij}\sigma_{cij}\rho_{bcij}}{\sqrt{1+ \sigma^2_{bij}}}\phi\left(\frac{\bm{X}^\top_{ij}\bm{\beta}}{\sqrt{1+\sigma^2_{bij}}}\right)\]
and
\[\Phi\left(\frac{\bm{X}^\top_{ij}\bm{\beta}}{\sqrt{1+\sigma^2_{bij}}}\right)\Delta_{ij}+\frac{\theta\sigma^2_{bij}}{\sqrt{1+ \sigma^2_{bij}}}\phi\left(\frac{\bm{X}^\top_{ij}\bm{\beta}}{\sqrt{1+ \sigma^2_{bij}}}\right)\]
respectively. The derivation of the first overall marginal mean of $g(Y_{ij})$ (resulting from models with two processes) can be found in the supplementary material of Tom {\it et al.},$^{5}$ and the second overall marginal mean of $g(Y_{ij})$ is derived in the appendix. If we specify $\mathbb{E}[g(Y_{ij})]=\bm{Z}^\top_{ij}\bm{\alpha}$, we can then reparametrise
\[\tag{22}\Delta_{ij}=\left[\bm{Z}^\top_{ij}\bm{\alpha}-\frac{\sigma_{bij}\sigma_{cij}\rho_{bcij}}{\sqrt{1+ \sigma^2_{bij}}}\phi\left(\frac{\bm{X}^\top_{ij}\bm{\beta}}{\sqrt{1+ \sigma^2_{bij}}}\right)\right]\Bigg{/}\Phi\left(\frac{\bm{X}^\top_{ij}\bm{\beta}}{\sqrt{1+\sigma^2_{bij}}}\right)\]
and
\[\Delta_{ij}=\left[\bm{Z}^\top_{ij}\bm{\alpha}-\frac{\theta\sigma^2_{bij}}{\sqrt{1+ \sigma^2_{bij}}}\phi\left(\frac{\bm{X}^\top_{ij}\bm{\beta}}{\sqrt{1+ \sigma^2_{bij}}}\right)\right]\Bigg{/}\Phi\left(\frac{\bm{X}^\top_{ij}\bm{\beta}}{\sqrt{1+\sigma^2_{bij}}}\right)\]
in the respective models. Thus, as in (20), $\bm{\alpha}$ offers easily interpretable covariate effects of $\bm{Z}_{ij}$ on the overall marginal mean of $g(Y_{ij})$ (by definition). In particular, a unit change in components of $\bm{Z}_{ij}$ will increase the overall marginal mean of $g(Y_{ij})$ by the respective components in $\bm{\alpha}$. However, from (21-22), it is also evident that replacing $\bm{Z}^\top_{ij}\bm{\gamma}$ with $\Delta_{ij}$ in Section 4 will still allow the proposed efficient estimation procedure to be applied. It is also possible to reparametrise patient-specific covariate effects $\bm{\beta}$ in the binary component in terms of population-level covariate effects $\bm{\xi}$, specifically $\bm{\beta}=\bm{\xi}\sqrt{1+\sigma^2_{bij}}$, since it can be shown that $\mathbb{P}(U_{ij}=1)=\Phi(\bm{X}^\top_{ij}\bm{\beta}/\sqrt{1+\sigma^2_{bij}})$. This relationship is easily proved. In the motivating application, this reparametrisation led to a numerically unstable optimization routine, therefore $(\bm{\beta},\bm{\alpha})$ was estimated with $\hat{\bm{\xi}}$ obtained as $\hat{\bm{\beta}}/\sqrt{1+\hat{\sigma}^2_{bij}}$ and standard errors were calculated using the delta method.\\

\subsection{Population-based inference}
Using the parameterisations described in the previous subsection, we demonstrate how population-based inference on the probability of being disabled and the overall marginal mean HAQ score can be obtained. Specifically, on averaging across patients, how a unit change in covariate values impacts these quantities. For illustrative purposes, the same covariates as those considered in the patient-specific case are considered. Note that for generalized linear models, conditional and marginal covariate effects will generally differ unless certain random effects distributions and link functions are chosen.$^{25}$\\
\indent As mentioned, marginal covariate effects on the probability of being disabled, $\bm{\xi}$, were obtained from $\bm{\beta}/\sqrt{1+\sigma^2_{bij}}$ with $\bm{\beta}$ and $\sigma^2_{bij}$ (the variance of $B_i(t_{ij})$) estimated using the model fitting procedure. The shared OU process and random effect models, where $\sigma^2_{bij}$ does not depend on $j$ (or $i$) for these models, were considered. For models with random walks, $\sigma^2_{bij}$ varies with $j$ and therefore $\bm{\xi}$ will have a time-dependent interpretation. For simplicity, these models are not considered. The shared random effect model was fitted using the parameterisation described by (20), with $g_1(y)=\Phi(y)$ and $g_2(y)=y$, and using the parameterisation described by (21-22), thus the same link functions are used and the inferences (at the population-level) from these models are comparable. These models will  be denoted as shared random effect model-overall and -conditional respectively. Note that unlike at the population-level, the patient-specific assumptions from the shared random effect model-overall and -conditional are vastly different. The shared random effect-overall model assumes that the overall patient-specific mean, i.e. $\mathbb{E}[Y_{ij}|B^r_i=b^r_i,C^r_i=c^r_i]$, has a linear form, namely $\bm{Z}^\top_{ij}\bm{\alpha}+c^r_i$. While the shared random effect-conditional model assumes that this quantity takes a particular non-linear form, namely $(\Delta_{ij}+c^r_i)\Phi(\bm{X}^\top_{ij}\bm{\beta}+b^r_i)$. As before, the shared random effect models (both -overall and -conditional) were fitted using numerical integration and maximum likelihood estimation under the assumption that the positive values follow a normal distribution with constant variance.\\
\indent Table 2 presents the results. Population-level covariate effects on the overall marginal mean are seen to be relatively similar across models due to the considerable overlap in confidence intervals. All three models are in strong agreement regarding the population-level covariate effect of the number of active joints. That is, on average, patients with an additional actively inflamed joint has an overall mean HAQ score increased by approximately 0.02. In contrast to the patient-specific case, the population-level covariate effects on the probability of being disabled are now more consistent across models. A generalized likelihood ratio test of $\rho_{gb}=1$ produced a $p$ value of $<0.001$ and therefore the shared OU process model is to be preferred over the shared random effect-conditional model. Log-likelihood values also indicate slight preference to the shared random effect-conditional model (-3507.63) over the shared random effect-overall model (-3582.78).
\begin{sidewaystable}
\caption{Table displaying population-level effects and corresponding $95\%$ Wald intervals on the probability of being disabled and the overall marginal mean HAQ score. *denotes the standardised version of the covariate.}
 \begin{center}
\begin{tabular}{l*{6}{c}c}
\toprule
&Shared OU process&Shared random effect-conditional&Shared random effect-overall\\
\midrule
Binary component &&\\
\midrule
Damaged joints &0.012 (0.006, 0.018)&0.0057 (0.0006, 0.011)&0.0066 (0.002, 0.011)\\
Active joints&0.05 (0.043, 0.058)&0.063 (0.054, 0.071) &0.051 (0.044, 0.058)\\
Sex &-0.65 (-0.88, -0.43) &-0.64 (-0.8, -0.48) &-0.61 (-0.76, -0.46)\\
Arthritis duration &0.013 (0.0057, 0.02) &0.014 (0.0095, 0.02)&0.012 (0.0078, 0.017)\\
Age at arthritis onset* &0.19 (0.14, 0.24) &0.2 (0.13, 0.27)&0.19 (0.12, 0.26)\\
Intercept &0.59 (0.35, 0.83) &0.56 (0.42, 0.7)&0.68 (0.55, 0.81) \\
\midrule
Overall marginal mean&&\\
\midrule
Damaged Joints &0.0064 (0.0036, 0.0094) &0.0029 (0.00054, 0.0053)&0.004 (0.0016, 0.0065)\\
Active joints  &0.02 (0.018, 0.022) &0.021 (0.019, 0.023)&0.022 (0.02, 0.024)\\
Sex &-0.3 (-0.4, -0.19) &-0.28 (-0.36, -0.21)&-0.31 (-0.39, -0.23)\\
Arthritis duration &0.0035 (0.00032, 0.0068) &0.006 (0.0038, 0.0082)&0.0054 (0.0031, 0.0077) \\
Age at arthritis onset* &0.073 (0.049, 0.098) &0.08 (0.048, 0.11)&0.09 (0.052, 0.13)\\
Intercept &0.62 (0.5, 0.73) &0.61 (0.54, 0.68)&0.59 (0.52, 0.66)\\
\midrule
$\theta$&0.18 (0.16, 0.2) &0.27 (0.24, 0.29)&0.22 (0.2, 0.24)\\
$\sigma^2$ &0.064 (0.057, 0.071) &0.12 (0.11, 0.12)&0.12 (0.11, 0.13)\\
$\sigma^2_b$ & &3.63 (2.92, 4.5)&5.19 (4.26, 6.34)\\
$\sigma^2_{gb}$  &11.12 (8.71, 14.19) &\\
$\rho_{gb}$ &0.95 (0.94, 0.96)  &\\
Log-likelihood &-3277.2 &-3507.63&-3582.78\\
\bottomrule
\end{tabular}
\end{center}
\end{sidewaystable}

\section{Discussion}
This paper reconsiders the flexible two-part models of Albert and Shen$^8$ and Ghosh and Albert,$^9$ and proposes an efficient method of implementation. Specifically, the problem of integrating over high dimensional random effects is replaced by evaluating the cumulative distribution function of a multivariate normal distribution. This leads to efficient algorithms being employed and results in only an optimization procedure being required for model fitting. Furthermore, while retaining the flexibility of including stochastic processes and the practicality of an efficient model fitting procedure, this paper also provides model parameterisations which allow easily interpretable covariate effects to act on the overall marginal mean. The proposed methodology was applied to a psoriatic arthritis data set with extensive follow-up information.\\
\indent Through their application and a simulation study, Albert and Shen$^8$ demonstrated that overall conditional means (conditional on realisations of stochastic processes) may suffer from bias if serial correlation is present but a shared random effect model is used instead. Furthermore, as the shared random effect model becomes more misspecified ($\rho_{gb}$ decreases from one) the degree of bias increases. However, under the same set-up, overall marginal means were less susceptible to bias. In the motivating application, the estimated degradation parameters from the shared OU process models were $\hat{\rho}_{gb}=0.95$ in both applications (Sections 5 and 6.1). The reasonably high estimated correlation may therefore explain why the shared random effect model $(\rho_{gb}=1)$  was a reasonable approximation in terms of estimating regression coefficients, although it was substantially the worst fitting model.\\
\indent Preliminary analyses suggested shared process models were reasonable for our data since $\hat{\rho}$, $\hat{\rho}_w$ and $\hat{\rho}_g\approx 1$ when the described bivariate processes models were fitted. Although this may not be surprising as both parts of the model are describing the same response process, it is worth noting that the estimated correlation parameter (between processes) can in principle take a value between $(-1,1)$ as evidenced in other works.$^{7,9}$ Our preliminary analyses also demonstrated the need for careful evaluation of models fitted as problems with over fitting may arise. This was evident when the estimated random variation parameter was estimated to be virtually zero (i.e. $\hat{\sigma}^2\approx0$) and the observed Fisher information matrix was non positive-definite, even when a considerably smaller tolerance level than the default was specified for computing multivariate normal probabilities.\\
\indent As mentioned in Section 6, the proposed model parameterisations were motivated by making inference on the overall marginal mean. In this regard, covariate effects (both patient-specific and population-level) on the mean of the positive values and its correlation structure were assumed not of interest. If the mean of the positive values is of primary interest, it would be more sensible to directly use (18-19), as in Section 5, to obtain patient-specific effects or derive a similar parameterisation, as in Section 6, to obtain population-level effects.\\
\indent A limitation of the current framework is that it is based on the assumption $g(Y_{ij})|Y_{ij}>0$ is approximately Gaussian with constant variance $\sigma^2$. Specifically, in situations where $g(\cdot)$ is required to be complex so that this assumption will at least approximately hold, the resulting inferential targets will no longer be intuitively interpretable owing to the complexity of the transformation function. One approach that may weaken the need to assume normality of $g(Y_{ij})|Y_{ij}>0$, particularly when the outcome exhibits a large amount of right skewness (e.g. medical expenditures), would be to make the alternative assumption $g(Y_{ij})|Y_{ij}>0$ follows a log-normal distribution. This may allow less complex and hence more interpretable transformation functions to be applied to the outcome without having to strongly violate the assumption on $g(Y_{ij})|Y_{ij}>0$. Under this alternative assumption, we provide details in the supplementary materials of how easily interpretable inference on the overall marginal mean and on the mean of the positive transformed outcomes can be obtained with computationally efficient likelihoods. Similar techniques to those in the supplementary materials can also be used when the assumption $g(Y_{ij})|Y_{ij}>0$ follows a log-skew-normal distribution is of interest. Although this comes at the cost of having an increased number of integrations in the marginal likelihood.\\
\indent Finally, the model described by (18-19) with possible simplifications described in Appendix B is very general. Although it was derived in the context of longitudinal semicontinuous data, it contains the model described by Barrett {\it et al.}$^{19}$ for the longitudinal and survival outcomes setting and implicitly provides a model for clustered cross-sectional semicontinuous data, where the index $(i,j)$ specifies the $j$th outcome from the $i$th cluster. The multivariate normal identity described in Section 4.2 can also facilitate the fitting of flexible models describing clustered binary data and continuous bounded outcome data$^{14}$. However, it should be noted that care is required when specifying an appropriate/suitable correlation structure. Particularly, the covariance matrix must be constrained to be symmetric and positive semi-definite otherwise the model fitting procedure will likely be problematic, as was found here. For these alternative situations, the proposed methodology does nevertheless offer a strong basis, especially with regard to implementation, for the developing of flexible models.

\begin{appendices}
\section{Rearranging (11)}
The continuous and random effects component in the integrand of (9) is
\[\left(\frac{1}{\sqrt{2\pi\sigma^2}}\right)^{\sum_{j=1}^{m_i}u_{ij}}\exp\left\{-\frac{(g(\bm{y}_i)-\bm{A}_{i3}\bm{\mu}_{ic}-\bm{A}_{i4}\bm{l}_i)^\top(g(\bm{y}_i)-\bm{A}_{i3}\bm{\mu}_{ic}-\bm{A}_{i4}\bm{l}_i)}{2\sigma^2}\right\}\phi^{\{\text{dim}(\bm{l}_i)\}}(\bm{l}_i;\bm{0},\bm{\Sigma}_{il}).\]
We rearrange this expression by completing the square in $\bm{l}_i$. This results in 
\begin{align*}\tag{23}
&\left(\frac{1}{\sqrt{2\pi\sigma^2}}\right)^{\sum_{j=1}^{m_i}u_{ij}}\exp\left\{-\frac{(g(\bm{y}_i)-\bm{A}_{i3}\bm{\mu}_{ic})^\top(g(\bm{y}_i)-\bm{A}_{i3}\bm{\mu}_{ic})}{2\sigma^2}\right\}\frac{|\bm{H}^{-1}_i|^{1/2}}{|\bm{\Sigma}_{il}|^{1/2}}\frac{1}{(2\pi)^{\text{dim}(\bm{l}_i)/2}|\bm{H}^{-1}_i|^{1/2}}\\
\times&\exp\left(-\frac{\bm{l}^\top_i\bm{H}_i\bm{l}_i}{2}+\bm{l}^\top_i\bm{H}_i\bm{h}_i\right).\\
\end{align*}
Focusing on terms containing $\bm{l}_i$, we have
\[\tag{24}\exp\left(-\frac{\bm{l}^\top_i\bm{H}_i\bm{l}_i}{2}+\bm{l}^\top_i\bm{H}_i\bm{h}_i\right)=\exp\left\{-\frac{(\bm{l}_i-\bm{h}_i)^\top\bm{H}_i(\bm{l}_i-\bm{h}_i)}{2}+\frac{\bm{h}^\top_i\bm{H}_i\bm{h}_i}{2}\right\}.\]
Equation (12-14) now follows by substituting (24) into (23).
\section{Simplification for correlated stochastic processes model}
For the models containing correlated stochastic processes (described by (7)), recall that $\bm{\Sigma}_{il}=\bm{\Sigma}_{i}$ (where $\bm{\Sigma}_i$ is described by either (5) or (6)), $\bm{A}_{i2}=(\text{diag}(2\bm{u}_i-{\bf{1}}),\text{diag}({\bf{0}}))$ and $\bm{A}_{i4}=(\text{diag}({\bf{0}}),\bm{u}_i)$ are $m_i\times 2m_i$ matrices. Simplification of (18) for this model structure is possible, specifically the following (to be derived) equations:
\begin{align*}\tag{25}
\bm{A}_{i2}\bm{h}_i&=\left(\bm{\Sigma}^u_{ibc}-\bm{\Sigma}^u_{ibc}(\bm{I}_{m_i}+\bm{\Sigma}^u_{ic}/\sigma^2)^{-1}\bm{\Sigma}^u_{ic}/\sigma^2\right)(g(\bm{y}_i)-\bm{A}_{i3}\bm{\mu}_{ic})/\sigma^2\\
\bm{A}_{i2}\bm{H}^{-1}_i\bm{A}^\top_{i2}&=\bm{\Sigma}^{u}_{ib}-\bm{\Sigma}^{u}_{ibc}(\bm{I}_{m_i}+\bm{\Sigma}^{u}_{ic}/\sigma^2)^{-1}\bm{\Sigma}^{u}_{ibc}/\sigma^2\\
|\bm{\Sigma}_i\bm{H}_i|&=|\bm{\Sigma}_{ic}\text{diag}(\bm{u}_i)/\sigma^2 +\bm{I}_{m_i}|\\
\bm{h}^\top_i\bm{H}_i\bm{h}_i&=(g(\bm{y}_i)-\bm{A}_3\bm{\mu}_{ic})^\top(\bm{\Sigma}^u_{ic}-\bm{\Sigma}^u_{ic}(\bm{I}_{m_i}+\bm{\Sigma}^u_{ic}/\sigma^2)^{-1}\bm{\Sigma}^u_{ic}/\sigma^2)(g(\bm{y}_i)-\bm{A}_{i3}\bm{\mu}_{ic})/\sigma^4
\end{align*}
where 
\begin{align*}
\bm{\Sigma}^u_{ib}&=\text{diag}(2\bm{u}_i-{\bf{1}})\bm{\Sigma}_{ib}\text{diag}(2\bm{u}_i-{\bf{1}})\\
\bm{\Sigma}^u_{ic}&=\text{diag}(\bm{u}_i)\bm{\Sigma}_{ic}\text{diag}(\bm{u}_i)\\
\bm{\Sigma}^u_{ibc}&=\text{diag}(2\bm{u}_i-{\bf{1}})\bm{\Sigma}_{ibc}\text{diag}(\bm{u}_i)
\end{align*}
reduce the dimension of the respective matrix calculations (from 2$m_i$ to $m_i$ dimensional).\\
In order to derive the equations in (25), we begin by simplifying $\bm{\Sigma}_i\bm{H}_i$. That is
\begin{align*}
\bm{\Sigma}_i\bm{H}_i&=\bm{\Sigma}_i\bm{A}^\top_{i4}\bm{A}_{i4}/\sigma^2+\bm{I}_{2m_i}\\
               &={\bordermatrix{&\cr
               &\bm{\Sigma}_{ib} &\bm{\Sigma}_{ibc}   \cr
               &\bm{\Sigma}_{ibc}&\bm{\Sigma}_{ic} \cr}}
{\bordermatrix{&\cr
               &\text{diag}({\bf{0}}) &\text{diag}({\bf{0}})   \cr
               &\text{diag}({\bf{0}})&\text{diag}(\bm{u}_i)/\sigma^2 \cr}}+\bm{I}_{2m_i}\\
&={\bordermatrix{&\cr
               &\bm{I}_{m_i} &\bm{\Sigma}_{ibc}\text{diag}(\bm{u}_i)/\sigma^2   \cr
               &\text{diag}({\bf{0}})&\bm{\Sigma}_{ic}\text{diag}(\bm{u}_i)/\sigma^2 +\bm{I}_{m_i}\cr}}.\\
\end{align*}
It now follows that
\[|\bm{\Sigma}_i\bm{H}_i|=|\bm{I}_{m_i}||\bm{\Sigma}_{ic}\text{diag}(\bm{u}_i)/\sigma^2 +\bm{I}_{m_i}|=|\bm{\Sigma}_{ic}\text{diag}(\bm{u}_i)/\sigma^2 +\bm{I}_{m_i}|.\]
Next, we simplify $\bm{h}^\top_i\bm{H}_i\bm{h}_i=(g(\bm{y}_i)-\bm{A}_{i3}\bm{\mu}_{ic})^\top\bm{A}_{i4}\bm{H}^{-1}_i\bm{A}^\top_{i4}(g(\bm{y}_i)-\bm{A}_{i3}\bm{\mu}_{ic})/\sigma^4$. By using the Woodbury matrix identity, specifically
\[\bm{H}^{-1}_i=\bm{\Sigma}_i-\bm{\Sigma}_i\bm{A}^\top_{i4}(\bm{I}_{m_i}+\bm{A}_{i4}\bm{\Sigma}_i\bm{A}^\top_{i4}/\sigma^2)^{-1}\bm{A}_{i4}\bm{\Sigma}_i/\sigma^2\]
and noting that
\[\bm{A}_{i4}\bm{\Sigma}_i\bm{A}^\top_{i4}=\text{diag}(\bm{u}_i)\bm{\Sigma}_{ic}\text{diag}(\bm{u}_i)\equiv\bm{\Sigma}^{u}_{ic},\]
we have
\begin{align*}
\bm{h}^\top_i\bm{H}_i\bm{h}_i&=(g(\bm{y}_i)-\bm{A}_{i3}\bm{\mu}_{ic})^\top\bm{A}_{i4}\bm{H}^{-1}\bm{A}^\top_{i4}(g(\bm{y}_i)-\bm{A}_{i3}\bm{\mu}_{ic})/\sigma^4\\
&=(g(\bm{y}_i)-\bm{A}_{i3}\bm{\mu}_{ic})^\top(\bm{\Sigma}^u_{ic}-\bm{\Sigma}^u_{ic}(\bm{I}_{m_i}+\bm{\Sigma}^u_{ic}/\sigma^2)^{-1}\bm{\Sigma}^u_{ic}/\sigma^2)(g(\bm{y}_i)-\bm{A}_{i3}\bm{\mu}_{ic})/\sigma^4.
\end{align*}
To simplify $\bm{A}_{i2}\bm{h}_i$, consider
\begin{align*}
\bm{A}_{i2}\bm{h}_i&=\bm{A}_{i2}\bm{H}^{-1}_i\bm{A}^\top_{i4}(g(\bm{y}_i)-\bm{A}_{i3}\bm{\mu}_{ic})/\sigma^2\\
&=\left(\bm{A}_{i2}\bm{\Sigma}_i\bm{A}^\top_{i4}-\bm{A}_{i2}\bm{\Sigma}_i\bm{A}^\top_{i4}(\bm{I}_{m_i}+\bm{\Sigma}^u_{ic}/\sigma^2)^{-1}\bm{\Sigma}^u_{ic}/\sigma^2\right)(g(\bm{y}_i)-\bm{A}_{i3}\bm{\mu}_{ic})/\sigma^2.
\end{align*}
By noting that
\[\bm{A}_{i2}\bm{\Sigma}_i\bm{A}^\top_{i4}=\text{diag}(2\bm{u}_i-{\bm{1}})\bm{\Sigma}_{ibc}\text{diag}(\bm{u}_i)\equiv\bm{\Sigma}^{u}_{ibc},\]
we have
\[\bm{A}_{i2}\bm{h}_i=\left(\bm{\Sigma}^u_{ibc}-\bm{\Sigma}^u_{ibc}(\bm{I}_{m_i}+\bm{\Sigma}^u_{ic}/\sigma^2)^{-1}\bm{\Sigma}^u_{ic}/\sigma^2\right)(g(\bm{y}_i)-\bm{A}_{i3}\bm{\mu}_{ic})/\sigma^2.\]
Furthermore
\[\bm{A}_{i2}\bm{H}^{-1}_i\bm{A}^\top_{i2}=\bm{\Sigma}^{u}_{ib}-\bm{\Sigma}^{u}_{ibc}(\bm{I}_{m_i}+\bm{\Sigma}^{u}_{ic}/\sigma^2)^{-1}\bm{\Sigma}^{u}_{ibc}/\sigma^2\]
where
\[\bm{\Sigma}^u_{ib}\equiv \bm{A}_{i2}\bm{\Sigma}_i\bm{A}^\top_{i2}=\text{diag}(2\bm{u}_i-\bm{1})\bm{\Sigma}_{ib}\text{diag}(2\bm{u}_i-\bm{1}).\]
\section{Overall marginal mean of shared process model}
To obtain the overall marginal mean of a shared process model, the following integral must be evaluated 
\begin{align*}\tag{26}
\int_b\Phi(\bm{X}^\top\bm{\beta}+b)(\bm{Z}^\top\bm{\gamma}+\theta b)\phi^{(1)}(b;0,\sigma^2_{b})\,db=&\bm{Z}^\top\bm{\gamma}\int_b\Phi(\bm{X}^\top\bm{\beta}+b)\phi^{(1)}(b;0,\sigma^2_{b})\,db\\
+&\theta\int_bb\Phi(\bm{X}^\top\bm{\beta}+b)\phi^{(1)}(b;0,\sigma^2_{b})\,db.
\end{align*}
The first term in (26) can be evaluated using the skew normal result, i.e. using (10), that is
\[\bm{Z}^\top\bm{\gamma}\int_b\Phi(\bm{X}^\top\bm{\beta}+b)\phi^{(1)}(b;0,\sigma^2_{b})\,db=\bm{Z}^\top\bm{\gamma}\Phi^{(1)}(\bm{X}^\top\bm{\beta};0,1+\sigma^2_b)=\bm{Z}^\top\bm{\gamma}\Phi\left(\frac{\bm{X}^\top\bm{\beta}}{\sqrt{1+\sigma^2_b}}\right).\]
To compute the second term in (26), consider
\[\theta\int_bb\Phi(\bm{X}^\top\bm{\beta}+b)\phi^{(1)}(b;0,\sigma^2_{b})\,db=\theta\sigma_b\int_{b^{*}}b^{*}\Phi(\bm{X}^\top\bm{\beta}+\sigma_bb^{*})\phi(b^{*})\,db^{*}.\]
This integral can be computed using equation (10, 011.3) in Owen,$^{26}$ which results in
\[\frac{\theta\sigma^2_{b}}{\sqrt{1+ \sigma^2_{b}}}\phi\left(\frac{\bm{X}^\top\bm{\beta}}{\sqrt{1+ \sigma^2_{b}}}\right).\]
The overall marginal mean now follows.
\end{appendices}

\section*{Acknowledgments}
We are grateful to Professor Vernon T. Farewell for providing general discussions on this research. We also acknowledge the patients in the Toronto Psoriatic Arthritis Clinic. This work was financially supported by the UK Medical Research Council [Unit program numbers U105261167 and MC\_UP\_1302/3].

\end{document}

\maketitle
\section*{Two-part log-normal model}
Another approach to modelling longitudinal semicontinuous data is to specify a two-part log-normal model for $Y_{ij}$. In particular, we assume that the model for the binary component $U_{ij}=I(Y_{ij}>0)$ is Bernoulli with response probability
\[\mathbb{P}(U_{ij}=1|b_{ij})=\Phi(\bm{X}^\top_{ij}\bm{\beta}+b_{ij})\]
and that the continuous component follows a log-normal distribution with probability density function
\[f(y_{ij}|Y_{ij}>0;c_{ij})=\frac{1}{y_{ij}\sqrt{2\pi\sigma^2}}\exp\left\{-\frac{(\log(y_{ij})-\bm{Z}^\top_{ij}\bm{\gamma}-c_{ij}+\sigma^2/2)^2}{2\sigma^2}\right\}\] 
where $\mathbb{E}[Y_{ij}|Y_{ij}>0;c_{ij}]=\exp(\bm{Z}^\top_{ij}\bm{\gamma}+c_{ij})$, $\sigma>0$ and $b_{ij}$ and $c_{ij}$ are realisations of the stochastic processes described in Sections 3.1 and 3.2 of the main text. Note that in contrast to the main text, the patient-specific mean of the positive values is now modelled using a log-link function and therefore a unit change of covariate values will multiply the patient-specific mean of the positive values by the exponent of the corresponding components in $\bm{\gamma}$.
\section*{Computationally efficient inference for the two-part log-normal model}
In this section, we demonstrate how computationally efficient inference for the two-part log-normal model can be obtained. Because the ideas are similar to Section 4 of the main text and consequently the derivations of the results follow in a similar manner, we omit minor details and state only the main results.\\
\indent For ease of exposition, we describe the likelihood contribution from patient $i$. The notation and conventions used in this section will be consistent with the main text. For models that contain two stochastic processes, the likelihood contribution from patient $i$ is
\begin{align*}\tag{A1}
L_i(\bm{\Theta})=&\int_{\bm{b_i}}\int_{\bm{c_i}}\left[\prod_{j=1}^{m_i}\Phi(\bm{X}^\top_{ij}\bm{\beta}+b_{ij})^{u_{ij}}\{1-\Phi(\bm{X}^\top_{ij}\bm{\beta}+b_{ij})\}^{1-u_{ij}}\right]\\
\times&\left[\prod_{j=1}^{m_i}\frac{1}{(y_{ij}+1-u_{ij})(\sqrt{2\pi\sigma^2})^{u_{ij}}}\exp\left\{-\frac{(\log(y_{ij}+1-u_{ij})-u_{ij}(\bm{Z}^\top_{ij}\bm{\gamma}+c_{ij}-\sigma^2/2))^2}{2\sigma^2}\right\}\right]\\
\times&\phi^{(2m_i)}(\bm{b}_i,\bm{c}_i;\bm{0},\bm{\Sigma}_i)\,d\bm{b}_i\,d\bm{c}_i.
\end{align*}
Similarly, for models containing a single stochastic process (i.e. shared process models), the likelihood contribution from patient $i$ is
\begin{align*}\tag{A2}
L_i(\bm{\Theta})=&\int_{\bm{b_i}}\left[\prod_{j=1}^{m_i}\Phi(\bm{X}^\top_{ij}\bm{\beta}+b_{ij})^{u_{ij}}\{1-\Phi(\bm{X}^\top_{ij}\bm{\beta}+b_{ij})\}^{1-u_{ij}}\right]\\
\times&\left[\prod_{j=1}^{m_i}\frac{1}{(y_{ij}+1-u_{ij})(\sqrt{2\pi\sigma^2})^{u_{ij}}}\exp\left\{-\frac{(\log(y_{ij}+1-u_{ij})-u_{ij}(\bm{Z}^\top_{ij}\bm{\gamma}+\theta b_{ij}-\sigma^2/2))^2}{2\sigma^2}\right\}\right]\\
\times&\phi^{(m_i)}(\bm{b}_i;\bm{0},\bm{\Sigma}_{ib})\,d\bm{b}_i.
\end{align*}
In matrix form, we describe a generic likelihood contribution that encapsulates (A1) and (A2):  
\begin{align*}\tag{A3}
&L_i(\bm{\Theta})=\int_{\bm{l}_i}\Phi^{(m_i)}\left(\bm{A}_{i1}\bm{\mu}_{ib}+\bm{A}_{i2}\bm{l}_i;\bm{0},\bm{I}_{m_i}\right)\left(\frac{1}{\sqrt{2\pi\sigma^2}}\right)^{\sum_{j=1}^{m_i}u_{ij}}\prod_{j=1}^{m_i}\frac{1}{y_{ij}+1-u_{ij}}\\
\times&\exp\left\{-\frac{(\log(\bm{y}_i+\bm{1}-\bm{u}_i)-\bm{A}_{i3}(\bm{\mu}_{ic}-\sigma^2/2)-\bm{A}_{i4}\bm{l}_i)^\top(\log(\bm{y}_i+\bm{1}-\bm{u}_i)-\bm{A}_{i3}(\bm{\mu}_{ic}-\sigma^2/2)-\bm{A}_{i4}\bm{l}_i)}{2\sigma^2}\right\}\\
\times&\phi^{\{\text{dim}(\bm{l}_i)\}}(\bm{l}_i;\bm{0},\bm{\Sigma}_{il})\,d\bm{l}_i
\end{align*}
where $\bm{l}_i=(\bm{b}_i,\bm{c}_i)$, $\bm{A}_{i2}=(diag(2\bm{u}_i-\bm{1}),diag(\bm{0}))$ and $\bm{A}_{i4}=(diag(\bm{0}),diag(\bm{u}_i))$ for (A1) and $\bm{l}_i=\bm{b}_i$, $\bm{A}_{i2}=diag(2\bm{u}_i-\bm{1})$ and $\bm{A}_{i4}=diag(\theta\bm{u}_i)$ for (A2). By noting the similarity of (A3) to (9) in the main text, particularly $g(\bm{y_i})$ and $\bm{\mu}_{ic}$ are replaced with $\log(\bm{y}_i+\bm{1}-\bm{u}_i)$ and $\bm{\mu}_{ic}-\sigma^2/2$ respectively and that there is an additional multiplicative constant $\prod_{j=1}^{m_i}1/(y_{ij}+1-u_{ij})$, it is clear that (A3) can be re-expressed as
\begin{align*}\tag{A4}
&L_i(\bm{\Theta})=\Phi^{(m_i)}\left(\bm{A}_{i1}\bm{\mu}_{ib}+\bm{A}_{i2}\bm{h}_i;\bm{0},\bm{I}_{m_i}+\bm{A}_{i2}\bm{H}^{-1}_i\bm{A}^\top_{i2}\right)\left(\frac{1}{\sqrt{2\pi\sigma^2}}\right)^{\sum_{j=1}^{m_i}u_{ij}}\prod_{j=1}^{m_i}\frac{1}{y_{ij}+1-u_{ij}}\frac{1}{|\bm{\Sigma}_{il}\bm{H}_i|^{1/2}}\\
\times&\exp\left\{-\frac{(\log(\bm{y}_i+\bm{1}-\bm{u}_i)-\bm{A}_{i3}(\bm{\mu}_{ic}-\sigma^2/2))^\top(\log(\bm{y}_i+\bm{1}-\bm{u}_i)-\bm{A}_{i3}(\bm{\mu}_{ic}-\sigma^2/2))}{2\sigma^2}+\frac{\bm{h}^\top_i\bm{H}_i\bm{h}_i}{2}\right\}
\end{align*}
where 
\begin{align*}\tag{A5}
\bm{A}_{i1}&=\text{diag}(2\bm{u}_i-{\bf{1}})\\
\bm{A}_{i3}&=\text{diag}(\bm{u}_i)\\
\bm{H}_i&=\bm{A}^\top_{i4}\bm{A}_{i4}/\sigma^2+(\bm{\Sigma}_{il})^{-1}\\
\bm{h}_i&=\bm{H}^{-1}_i\bm{A}^\top_{i4}(\log(\bm{y}_i+\bm{1}-\bm{u}_i)-\bm{A}_{i3}(\bm{\mu}_{ic}-\sigma^2/2))/\sigma^2
\end{align*}
and $\bm{\Sigma}_{il}=\bm{\Sigma}_i$ or $\bm{\Sigma}_{ib}$ with $\bm{A}_{i2}$ and $\bm{A}_{i4}$ defined by the specified model. Because the cumulative distribution function of the multivariate normal distribution is computationally efficient to evaluate, (A4) provides an appealing approach to evaluate the likelihood contributions described by (A1) and (A2). \\ 
\section*{Modelling the overall marginal mean with the two-part log-normal model}
In order to directly model the overall marginal mean, i.e. $\mathbb{E}[Y_{ij}]$, using the two-part log-normal model, we follow the approach of Section 6 by reparameterising the mean of the positive values, i.e. $\mathbb{E}[Y_{ij}|Y_{ij}>0]$, whilst retaining the same random effects structure that facilitates the computationally efficient implementation procedure. Specifically, following (21) in the main text, let
\begin{align*}
\mathbb{P}(U_{ij}=1|B_i(t_{ij})=b_{ij})&=\Phi(\bm{X}^\top_{ij}\bm{\beta}+b_{ij})\\
\mathbb{E}[Y_{ij}|Y_{ij}>0,C_i(t_{ij})=c_{ij}]&=\exp(\Delta_{ij}+c_{ij})\\
\end{align*}
where $\Delta_{ij}$ is a function of covariates (at the $j$th visit from patient $i$) and regression coefficients only. The overall marginal mean for the correlated processes and shared process model is given by
\[\mathbb{E}[Y_{ij}]=\int_{-\infty}^{\infty}\int_{-\infty}^{\infty}\Phi(\bm{X}^\top_{ij}\bm{\beta}+b_{ij})\exp(\Delta_{ij}+c_{ij})\phi^{(2)}(b_{ij},c_{ij};\bm{0},\bm{\Sigma}_{ij})\,db_{ij}\,dc_{ij}\]
and
\[\mathbb{E}[Y_{ij}]=\int_{-\infty}^{\infty}\Phi(\bm{X}^\top_{ij}\bm{\beta}+b_{ij})\exp(\Delta_{ij}+\theta b_{ij})\phi^{(1)}(b_{ij};0,\sigma^2_{bij})\,db_{ij}\]
respectively. We begin by computing the overall marginal mean for the shared process model. Firstly, we have
\begin{align*}
\mathbb{E}[Y_{ij}]=&\int_{-\infty}^{\infty}\Phi(\bm{X}^\top_{ij}\bm{\beta}+b_{ij})\exp(\Delta_{ij}+\theta b_{ij})\phi^{(1)}(b_{ij};0,\sigma^2_{bij})\,db_{ij}\\
=&\exp(\Delta_{ij})\int_{-\infty}^{\infty}\int_{-\infty}^{\bm{X}^\top_{ij}\bm{\beta}}\phi(u+b_{ij})\,du\exp(\theta b_{ij})\phi^{(1)}(b_{ij};0,\sigma^2_{bij})\,db_{ij}.\\
\end{align*}
Next, we change the order of integration and complete the square in $b_{ij}$. This results in
\[\frac{\exp(\Delta_{ij})}{2\pi\sigma_{bij}}\int_{-\infty}^{\bm{X}^\top_{ij}\bm{\beta}}\int_{-\infty}^{\infty}\exp\left\{-\frac{1+\sigma^2_{bij}}{2\sigma^2_{bij}}\left(b+\frac{\sigma^2_{bij}(u-\theta)}{1+\sigma^2_{bij}}\right)^2\right\}\exp\left\{-\frac{u^2}{2}+\frac{\sigma^2_{bij}(u-\theta)^2}{2(1+\sigma^2_{bij})}\right\}\,db_{ij}\,du.\]
After evaluating the above integral with respect to $b_{ij}$ and then completing the square in $u$, we have
\[\mathbb{E}[Y_{ij}]=\frac{\exp\left\{\Delta_{ij}+\theta^2\sigma^2_{bij}(1+\theta^2\sigma^2_{bij})/\{2(1+\sigma^2_{bij})\}\right\}}{\sqrt{2\pi(1+\sigma^2_{bij})}}\int_{-\infty}^{\bm{X}^\top_{ij}\bm{\beta}}\exp\left\{-\frac{(u+\theta\sigma^2_{bij})^2}{2(1+\sigma^2_{bij})}\right\}\,du.\]
The overall marginal mean then results by making the substitution $v=(u+\theta\sigma^2_{bij})/\sqrt{1+\sigma^2_{bij}}$. Specifically,
\begin{align*}
\mathbb{E}[Y_{ij}]=&\exp\left\{\Delta_{ij}+\frac{\theta^2\sigma^2_{bij}(1+\theta^2\sigma^2_{bij})}{2(1+\sigma^2_{bij})}\right\}\int_{-\infty}^{\frac{\bm{X}^\top_{ij}\bm{\beta}+\theta\sigma^2_{bij}}{\sqrt{1+\sigma^2_{bij}}}}\phi(v)\,dv\\
=&\exp\left\{\Delta_{ij}+\frac{\theta^2\sigma^2_{bij}(1+\theta^2\sigma^2_{bij})}{2(1+\sigma^2_{bij})}\right\}\Phi\left(\frac{\bm{X}^\top_{ij}\bm{\beta}+\theta\sigma^2_{bij}}{\sqrt{1+\sigma^2_{bij}}}\right)
\end{align*}
The overall marginal mean for the correlated processes model can be computed as follows:
\begin{align*}
\mathbb{E}[Y_{ij}]=&\int_{-\infty}^{\infty}\int_{-\infty}^{\infty}\Phi(\bm{X}^\top_{ij}\bm{\beta}+b_{ij})\exp(\Delta_{ij}+c_{ij})\phi^{(2)}(b_{ij},c_{ij};\bm{0},\bm{\Sigma}_{ij})\,db_{ij}\,dc_{ij}\\
=&\int_{-\infty}^{\infty}\int_{-\infty}^{\infty}\Phi(\bm{X}^\top_{ij}\bm{\beta}+b_{ij})\exp(\Delta_{ij}+c_{ij})\phi^{(1)}(b_{ij};0,\sigma^2_{bij})\\
\times&\phi^{(1)}\left(c_{ij};\frac{\sigma_{cij}\rho_{bcij}b_{ij}}{\sigma_{bij}},(1-\rho^2_{bcij})\sigma^2_{cij}\right)\,db_{ij}\,dc_{ij}\\
=&\int_{-\infty}^{\infty}\Phi(\bm{X}^\top_{ij}\bm{\beta}+b_{ij})\phi^{(1)}(b_{ij};0,\sigma^2_{bij})\int_{-\infty}^{\infty}\exp(\Delta_{ij}+c_{ij})\\
&\times\phi^{(1)}\left(c_{ij};\frac{\sigma_{cij}\rho_{bcij}b_{ij}}{\sigma_{bij}},(1-\rho^2_{bcij})\sigma^2_{cij}\right)\,dc_{ij}\,db_{ij}\\
=&\exp\left\{\frac{(1-\rho^2_{bcij})\sigma^2_{cij}}{2}\right\}\int_{-\infty}^{\infty}\Phi(\bm{X}^\top_{ij}\bm{\beta}+b_{ij})\exp\left(\Delta_{ij}+\frac{\sigma_{cij}\rho_{bcij}b_{ij}}{\sigma_{bij}}\right)\phi^{(1)}(b_{ij};0,\sigma^2_{bij})\,db_{ij}
\end{align*}
where the second and third line results from factorizing the bivariate normal density into a conditional and marginal density and the sixth line results from using the moment generating function of the normal distribution. Finally, the overall marginal mean for the correlated processes model can be obtained by noting that the above integral is equivalent to the overall marginal mean of the shared process model with $\theta=\sigma_{cij}\rho_{bcij}/\sigma_{bij}$. Thus we have
\begin{align*}
\mathbb{E}[Y_{ij}]=&\exp\left\{\frac{(1-\rho^2_{bcij})\sigma^2_{cij}}{2}\right\}\exp\left\{\Delta_{ij}+\frac{\rho^2_{bcij}\sigma^2_{cij}(1+\rho^2_{bcij}\sigma^2_{cij})}{2(1+\sigma^2_{bij})}\right\}\Phi\left(\frac{\bm{X}^\top_{ij}\bm{\beta}+\rho_{bcij}\sigma_{bij}\sigma_{cij}}{\sqrt{1+\sigma^2_{bij}}}\right)\\
=&\exp\left\{\Delta_{ij}+\frac{\sigma^2_{cij}}{2}\left(1-\frac{\rho^2_{bcij}(\sigma^2_{bij}-\rho^2_{bcij}\sigma^2_{cij})}{1+\sigma^2_{bij}}\right)\right\}\Phi\left(\frac{\bm{X}^\top_{ij}\bm{\beta}+\rho_{bcij}\sigma_{bij}\sigma_{cij}}{\sqrt{1+\sigma^2_{bij}}}\right).
\end{align*}
It is now implicit that replacing $\Delta_{ij}$ with
\[\bm{Z}^\top_{ij}\bm{\alpha}-\frac{\theta^2\sigma^2_{bij}(1+\theta^2\sigma^2_{bij})}{2(1+\sigma^2_{bij})}-\log\left\{\Phi\left(\frac{\bm{X}^\top_{ij}\bm{\beta}+\theta\sigma^2_{bij}}{\sqrt{1+\sigma^2_{bij}}}\right)\right\}\]
and 
\[\bm{Z}^\top_{ij}\bm{\alpha}-\frac{\sigma^2_{cij}}{2}\left(1-\frac{\rho^2_{bcij}(\sigma^2_{bij}-\rho^2_{bcij}\sigma^2_{cij})}{1+\sigma^2_{bij}}\right)-\log\left\{\Phi\left(\frac{\bm{X}^\top_{ij}\bm{\beta}+\rho_{bcij}\sigma_{bij}\sigma_{cij}}{\sqrt{1+\sigma^2_{bij}}}\right)\right\}\]
in the shared process and correlated processes models respectively will results in $\mathbb{E}[Y_{ij}]=\exp(\bm{Z}^\top_{ij}\bm{\alpha})$. Thus the proposed parameterisations allow easily interpretable covariate effects to act on the overall marginal mean, particularly a unit change in covariate values multiplies the overall marginal mean by the exponent of the respective components in $\bm{\alpha}$, while retaining the proposed computationally efficient implementation procedure.